\documentclass[12pt]{article}
\usepackage{amsmath}
\usepackage{amssymb}
\begin{document}
%
%
\baselineskip=18pt
\begin{titlepage}
%
%
%
\begin{flushright}
KOBE-TH-09-04\\
\end{flushright}
\vspace{1cm}
\begin{center}{\Large\bf 
Strong Coupling Quantum Einstein Gravity\\
at a $\boldsymbol{z=2}$ Lifshitz Point
}
\end{center}
\vspace{1cm}
\begin{center}
Makoto Sakamoto$^{(a)}$
\footnote{E-mail: dragon@kobe-u.ac.jp}
\end{center}
\vspace{0.2cm}
\begin{center}
${}^{(a)}$ {\it Department of Physics, Kobe University, 
Rokkodai, Nada, Kobe 657-8501, Japan}
\\[0.2cm]
\end{center}
\vspace{1cm}
\begin{abstract}
We solve a renormalized Wheeler-DeWitt equation for Einstein gravity
in $D+1$ dimensions with $D=$ odd
in the strong coupling limit,
which is expected to be suited to probe quantum geometry at
short distances,
in order to test Ho\v{r}ava's idea
that quantum gravity at short distances will be described by
a nonrelativistic system with dynamical critical exponent $z>1$.
Our results support the idea and show that the Wheeler-DeWitt
equation  possesses a solution associated with a $z=2$
Lifshitz point but no other $z>2$ solutions to leading order
of the strong coupling expansion.
\end{abstract}
\end{titlepage}
%
%
%

\newpage
%
%
%
\section{Introduction}
Ho\v{r}ava \cite{HoravaGR1,HoravaGR2} has recently
proposed an interesting idea that quantum gravity at short
distances will be described by a nonrelativistic system with
an anisotropy between space and time.
The anisotropy is characterized by the dynamical critical
exponent $z$.
A number of studies have been recently made on the
Ho\v{r}ava-Lifshitz gravity to examine its cosmological
applications \cite{cosmology}, black hole physics \cite{BH},
theoretical aspects \cite{theory}, etc.
However, little attention has been paid to a possibility that
quantum theory of Einstein general relativity will have 
a $z\!>\!1$ Lifshitz point at short distances, or that
quantum Einstein gravity will be described effectively
by a nonrelativistic theory at high energies, where
Lorentz invariance should be broken spontaneously 
in the UV\footnote{
An attempt has been made in \cite{Moffat}.
}
rather than emerging accidentally in the infra-red.
This is the perspective we adopt in this paper.

In \cite{SakamotoGR1,SakamotoGR2}, we have presented a 
renormalization  prescription of the $3+1$-dimensional
Wheeler-DeWitt equation \cite{WD}, and
found a solution in the strong coupling expansion, which
is expected to probe short distance behavior of quantum
gravity \cite{strong_coupling}.
Since the formalism is particularly suited to
test Ho\v{r}ava's idea, it will be worth while reinvestigating
the previous study of the Wheeler-DeWitt equation from
a new aspect of Ho\v{r}ava's proposal, and extending
the analysis to higher dimensions.
The purpose of this paper is to answer the question whether
the renormalized Wheeler-DeWitt equation for Einstein gravity
in $D+1$ dimensions can possess any solutions associated 
with $z\!>\!1$ Lifshitz points to leading order of 
the strong coupling expansion.
Our answer is positive and we find a $z\!=\!2$ solution
but no other $z\!>\!2$ solutions for $D=$ odd.


This paper is organized as follows.
In section $2$, we explain our renormalization prescription
for the Wheeler-DeWitt equation in $D+1$ dimensions.
In section $3$, we discuss consistency of the constraints with
our renormalization procedure.
In section $4$, we solve the renormalized Wheeler-DeWitt
equation to leading order of the strong coupling expansion.
Section $5$ is devoted to conclusions.

\vspace{5mm}
%
%
%
\section{Renormalization Scheme}
In this section, we present a renormalization prescription for
the Wheeler-DeWitt equation in $D+1$ dimensions.
The prescription has been developed by Mansfield \cite{Mansfield}
for the functional Schr\"{o}dinger equation of the Yang-Mills
theory and then generalized to the Wheeler-DeWitt 
equation \cite{SakamotoGR1,SakamotoGR2}.
In the following, we extend the technique given in 
\cite{SakamotoGR1,SakamotoGR2}
to arbitrary dimensions.
Since actual computations are tedious and
lengthy, we will omit the details in this paper.
Instead, we will refer the reader to the appendices of
\cite{SakamotoGR2} for $D=3$.
We will ignore boundary terms and freely drop integrals of
all total derivative terms.
The inclusion of those effects is beyond the scope of the paper.

The (unregulated) Wheeler-DeWitt equation \cite{WD} is given by
%
\begin{eqnarray}
\left[\, - 16\pi G\, G_{ijkl}(x) \frac{\delta}{\delta g_{kl}(x)}
 \frac{\delta}{\delta g_{ij}(x)}
  + \frac{\sqrt{g(x)}}{16\pi G} 
   \Big( R(x) + 2\Lambda \Big) \,\right] \Psi[g] = 0\,,
\label{eq2.1}
\end{eqnarray}
%
where $G$ is the \lq\lq Newton\rq\rq \,constant in $D+1$ 
dimensions and $G_{ijkl}(x)$ is the metric on superspace
%
\begin{eqnarray}
G_{ijkl}(x) = \frac{1}{2\sqrt{g(x)}}\Big(
  g_{ik}(x)g_{jl}(x) + g_{il}(x)g_{jk}(x) 
   - \frac{2}{D-1}\, g_{ij}(x)g_{kl}(x)\Big) \,.
\label{eq2.2}
\end{eqnarray}
%
The $R(x)$ denotes the scalar curvature constructed from the
$D$-dimensional metric $g_{ij}(x)$ and $\Lambda$ is a
cosmological constant.
The equation (\ref{eq2.1}) is ill defined because a product
of two functional derivatives at the same point
%
\begin{eqnarray}
\Delta(x) \equiv G_{ijkl}(x) \frac{\delta}{\delta g_{kl}(x)}
 \frac{\delta}{\delta g_{ij}(x)} 
\label{eq2.3}
\end{eqnarray}
%
could produce the square of $\delta$-functions,
like $(\delta^{D}(x,x'))^2$, which is meaningless.
To make (\ref{eq2.3}) well defined, we need to replace
$\Delta(x)$ by a renormalized operator $\Delta_{R}(x)$.
To this end, we first define a regularized differential
operator $\Delta(x;s)$ by point-splitting the functional
derivatives by use of a heat kernel \cite{SakamotoGR1,SakamotoGR2}:
%
\begin{eqnarray}
\Delta(x\,;s) \equiv 
  \int \!d^{D}\!x'\ K_{i'j'kl}(x',x\,;s) \frac{\delta}{\delta g_{kl}(x)}
 \frac{\delta}{\delta g_{i'j'}(x')}\,, 
\label{eq2.4}
\end{eqnarray}
%
where $K_{i'j'kl}(x',x\,;s)$ is a bitensor at both $x'$
and $x$ and satisfies the heat equation
%
\begin{eqnarray}
-\frac{\partial}{\partial s}\,K_{i'j'kl}(x',x\,;s) 
 = - \nabla^{\prime}_{p}\nabla^{\prime p}\, K_{i'j'kl}(x',x\,;s) 
\label{eq2.5}
\end{eqnarray}
%
with the initial condition
%
\begin{eqnarray}
\lim_{s\,\to\, 0} K_{i'j'kl}(x',x\,;s) 
 = G_{i'j'kl}(x)\,\delta^{D}(x',x)\,, 
\label{eq2.6}
\end{eqnarray}
%
which assures that $\Delta(x\,;s)$ reduces to $\Delta(x)$ in the
naive limit $s \to 0$.
Here, $\nabla^{\prime}_{p}$ denotes the covariant derivative
with respect to $x'$.
Taking $s$ small but nonzero in (\ref{eq2.4}) gives a regularized
operator of $\Delta(x)$.
We have chosen the factor ordering written in (\ref{eq2.4}).
Other choices of factor ordering will lead to different numerical
values of our results but will not change the qualitative features.

The heat equation (\ref{eq2.5}) can be solved by the standard
technique \cite{Schwinger-DeWitt, dynamical theory}.
Let $\cal{O}$ be $D$-dimensional integrals of local functions
of $g_{ij}$.
The action of $\Delta(x\,;s)$ on $\cal{O}$ will give an expansion
in powers of $s$.
These powers of $s$ may be determined from general coordinate
invariance and dimensional analysis.
For example, we have
%
\begin{eqnarray}
\Delta(x\,;s) \!\!\int\! d^{D}\!y\sqrt{g(y)}
 \!\!\!&=&\!\!\! \frac{\sqrt{g(x)}}{s^{D/2}}
   \left\{ \alpha_{0}
    + \sum_{a}\sum_{n=1}^{\infty} 
     s^{n}\,\alpha^{a}_{n} {\cal{O}}^{a}_{2n}(x)
   \right\},
\label{eq2.7}\\
\Delta(x\,;s) \!\!\int\! d^{D}\!y\sqrt{g(y)} R(y)
 \!\!\!&=&\!\!\! \frac{\sqrt{g(x)}}{s^{D/2+1}}
   \left\{ \beta_{0}
    + s\,\beta_{1} R(x)
    + \sum_{a}\sum_{n=1}^{\infty} 
      s^{n+1}\beta^{a}_{n+1} {\cal{O}}^{a}_{2(n+1)}(x)
   \right\},
\label{eq2.8}
\end{eqnarray}
%
where $\{{\cal O}^{a}_{2n}(x)\}$ symbolically denote a set of independent
local scalar functions of mass dimension $2n$, and $\alpha^{a}_{n},
\beta^{a}_{n}$ are dimensionless numerical constants.
The first few coefficients are found to be
%
\begin{eqnarray}
\alpha_{0}
 \!\!&=&\!\! -\,\frac{D(D^{2}-2)}{4(D-1)(4\pi)^{D/2}}\,,\nonumber\\
\beta_{0}
 \!\!&=&\!\! \,\frac{(D-2)D(D+1)}{8(4\pi)^{D/2}}\,,\nonumber\\
\beta_{1}
 \!\!&=&\!\! -\,\frac{(D-2)(D^{3}+10D^{2}-13D-34)}{48(D-1)(4\pi)^{D/2}}\,.
\label{eq2.9}
\end{eqnarray}
%
These results agree with the values given 
in \cite{SakamotoGR1,SakamotoGR2} for $D=3$.
Note that $\beta_{0}=\beta_{1}=0$ for $D=2$.
This is consistent with the fact that the term
$\int d^{D}\!y \sqrt{g} R$ is a topological invariant for $D=2$.

The second step of our renormalization prescription is to extract
a finite part from $\Delta(x\,;s) {\cal O}$.
We define $\Delta_{R}(x) {\cal O}$ from $\Delta(x\,;s) {\cal O}$
by analytic continuation \cite{Mansfield}:
%
\begin{eqnarray}
\Delta_{R}(x) {\cal O}
 \equiv \lim_{s\,\to\,+0} s\int_{0}^{\infty} \!d\varepsilon\,
  \varepsilon^{s-1} \phi(\varepsilon)\, \Delta(x\,;s=\varepsilon^{2}) {\cal O}\,.
\label{eq2.10}
\end{eqnarray}
%
Note that if $\Delta(x\,;s=0){\cal O}$ is finite, 
$\Delta_{R}(x) {\cal O}$ reduces to  $\Delta(x\,;s=0){\cal O}$
provided that the differentiable function $\phi(\varepsilon)$
rapidly decreases to zero at infinity with 
%
\begin{eqnarray}
\phi(0) = 1\,.
\label{eq2.11}
\end{eqnarray}
%
According to the above renormalization prescription, we have,
for example,
%
\begin{eqnarray}
\Delta_{R}(x) \!\!\int\! d^{D}\!y\sqrt{g(y)}
  \!\!&=&\!\! \sqrt{g(x)}
   \left\{ \frac{\phi^{(D)}(0)}{D!}\,\alpha_{0}
    + \sum_{a}\sum_{n=1}^{[D/2]} 
     \frac{\phi^{(D-2n)}(0)}{(D-2n)!}\,\alpha^{a}_{n}\, 
     {\cal{O}}^{a}_{2n}(x)
   \right\},\ \ \ \ 
\label{eq2.12}\\
\Delta_{R}(x) \!\!\int\! d^{D}\!y\sqrt{g(y)} R(y)
  \!\!&=&\!\! \sqrt{g(x)}\,
   \Bigg\{ \frac{\phi^{(D+2)}(0)}{(D+2)!}\beta_{0}
    +\frac{\phi^{(D)}(0)}{D!}\beta_{1}R(x)\nonumber\\
&&\qquad\qquad + \sum_{a}\sum_{n=1}^{[D/2]} 
     \frac{\phi^{(D-2n)}(0)}{(D-2n)!}\,\beta^{a}_{n+1}\, 
     {\cal{O}}^{a}_{2(n+1)}(x)
   \Bigg\}\,,
\label{eq2.13}
\end{eqnarray}
%
where $\phi^{(n)}(0) \equiv d^{n}\phi(0)/d\varepsilon^{n}$, and
$[D/2]$ denotes the Gauss symbol ($[D/2]=(D-1)/2$ for $D=$ odd,
$[D/2]=D/2$ for $D=$ even).
The results depend on the arbitrary function $\phi$.
This is an inevitable consequence of isolating finite quantities 
from divergent ones. 
Physical quantities must be independent of this arbitrariness, 
so that coupling \lq\lq constants"
should be regarded as functions of $\phi$. 
This is the basic problem of renormalization
\cite{Mansfield,SakamotoGR1,SakamotoGR2}.
We will return to this point later.

\vspace{5mm}
%
%
%
\section{Consistency of Constraints}
We have chosen the renormalization prescription to preserve
$D$-dimensional general coordinate invariance.
This does not, however, guarantee the whole symmetry
of the theory at quantum level.
We have to ensure that our renormalization procedure
is consistent with the constraints which are generators
of the symmetry.

The constraints consist of the momentum constraint 
${\cal H}_{i}(x)$ and the  Hamiltonian constraint
${\cal H}(x)$.
Since our renormalization procedure preserves $D$-dimensional
general coordinate invariance, no anomalous terms may
appear in commutators with the momentum operators.
In our renormalization prescription, ${\cal H}(x)$
should be replaced by the renormalized Hamiltonian
constraint
%
\begin{eqnarray}
{\cal H}_{R}(x) \equiv
 -16\pi G\,\Delta_{R}(x) + \frac{\sqrt{g(x)}}{16\pi G}
  \,\Big( R(x) + 2\Lambda \Big).
\label{eq3.1}
\end{eqnarray}
%
Anomalous terms would appear in the commutator of ${\cal H}_{R}$'s :
%
\begin{eqnarray}
&&\Big[ \int \! d^{D}\!x\, \eta_{1}(x)\,{\cal H}_{R}(x)\,,
 \int \! d^{D}\!y\, \eta_{2}(y)\,{\cal H}_{R}(y) \Big] \nonumber\\
&&\qquad\qquad
 = i \int\! d^{D}\!x \Big( \eta_{1}(x)\big(\nabla_{i}\eta_{2}(x)\big)
   - \big(\nabla_{i}\eta_{1}(x)\big) \eta_{2}(x)\Big) {\cal H}^{i}(x)
   + \Delta \Gamma , 
\label{eq3.2}
\end{eqnarray}
%
where $\eta_{1}$ and $\eta_{2}$ are arbitrary scalar functions.
The anomalous term $\Delta\Gamma$ is expected to be of the form
%
\begin{eqnarray}
\Delta\Gamma = 
 \int\! d^{D}\!x\sqrt{g(x)} \sum_{a}\sum_{n=1}^{[D/2]}
  \frac{\phi^{(D-2n)}(0)}{(D-2n)!}\,\tilde{\cal O}_{2(n+1)}
   (\eta_{1},\eta_{2}\,;x)\,. 
\label{eq3.3}
\end{eqnarray}
%
We notice that dimension zero and two operators 
$\tilde{\cal O}_{0}$ and $\tilde{\cal O}_{2}$ will not
appear on the right-hand-side of (\ref{eq3.3}) because
there are no such operators satisfying the antisymmetry
under the exchange of $\eta_{1}$ and $\eta_{2}$.
There exists a dimension four operator $\tilde{\cal O}_{4}$,
which is the lowest operator satisfying the antisymmetric
property, i.e.
$\tilde{\cal O}_{4} =
 ( \eta_{1}( \nabla_{i}\,\eta_{2}) 
  - (\nabla_{i}\,\eta_{1}) \eta_{2}) \nabla^{i}R$,
up to a constant.
%
In \cite{SakamotoGR1,SakamotoGR2}, it has been shown that 
$\Delta\Gamma$ is
proportional to $\phi^{(1)}(0) \tilde{\cal O}_{4}$ with a
nonzero coefficient for $D=3$, as expected in (\ref{eq3.3}). 

We could, in principle, compute the right-hand-side of 
(\ref{eq3.3}) but it is practically impossible for any $D$
because the size of computations will increase rapidly with $D$.
It will not, however, be unreasonable to assume that every 
$\tilde{\cal O}_{2(n+1)}$ is non-vanishing for 
$n = 1,2,\cdots, [D/2]$ in (\ref{eq3.3}) even for $D>3$
since there is no symmetry to prevent every term in (\ref{eq3.3})
from appearing.
Thus, we conclude that the anomaly free condition 
$\Delta\Gamma = 0$ requires that
%
\begin{eqnarray}
\phi^{(D-2)}(0) = \phi^{(D-4)}(0) = \cdots =\phi^{(1)}(0) = 0 
\label{eq3.6}
\end{eqnarray}
%
for $D=$ odd.
For $D=$ even, there is an obstacle.
If $\tilde{\cal O}_{D+2}$ is nonzero, $\Delta\Gamma$ cannot be zero
for $D=$ even because of the condition (\ref{eq2.11}).
We will hereafter restrict ourselves to the case of $D=$ odd,
unless stated otherwise.

\vspace{5mm}
%
%
%
\section{
Strong Coupling Expansion and $\boldsymbol{z\!>\!1}$ Lifshitz Points
}
Let us now solve the renormalized Wheeler-DeWitt equation
%
\begin{eqnarray}
\left[\, - 16\pi G\, \Delta_{R}(x)  + \frac{\sqrt{g(x)}}{16\pi G} 
   \Big( R(x) + 2\Lambda \Big) \right] \Psi[g] = 0
\label{eq4.1}
\end{eqnarray}
%
with the anomaly free condition (\ref{eq3.6}).
Since we are interested in short distance behavior,
we do not probably need to solve it exactly.
The strong coupling expansion will be well suited to probe
quantum geometry at short distances 
\cite{strong_coupling,SakamotoGR1,SakamotoGR2}.
We then assume that the wave functional $\Psi[g]$ has the form
%
\begin{eqnarray}
\Psi[g] 
 \equiv \exp\Big\{-S[g]\,\Big\}
 = \exp \left\{\, -\sum_{n=1}^{\infty} \left( \frac{1}{16\pi G}
    \right)^{2n} \,S_{n}[g]\,\right\}\,.
\label{eq4.2}
\end{eqnarray}
%
Substituting (\ref{eq4.2}) into (\ref{eq4.1}),
we have the leading order equation
%
\begin{eqnarray}
\Delta_{R}(x)\, S_{1}[g] 
 = -\sqrt{g(x)}\,\big( R(x) + 2\Lambda\,\big)\,.
\label{eq4.4}
\end{eqnarray}
%
Since we are looking for solutions that realize Ho\v{r}ava's idea
\cite{HoravaGR1,HoravaGR2}, we restrict the form of $S_{1}[g]$
to a finite sum of integrals of local functions of $g_{ij}$:
%
\begin{eqnarray}
S_{1}[g] = \int\! d^{D}\!x\,\sqrt{g(x)}\sum_{a}\sum_{n=0}^{N}
  \,\gamma^{a}_{n}\,{\cal O}^{a}_{2n}(x)\,,
\label{eq4.5}
\end{eqnarray}
%
where $\gamma^{a}_{n}$ is a dimensionful constant of mass
dimension $-D-2n+2$.
According to the discussions given in \cite{HoravaGR1,HoravaGR2},
the solution (\ref{eq4.5}) turns out to realize 
a $z=2N$ Lifshitz point 
if one of the coefficients $\gamma^{a}_{N}$ for the highest 
operators given by the form ${\cal O}^{a}_{2N} \sim (\nabla)^{2N-2}R$ 
is non-vanishing.

It is easy to see that the relations (\ref{eq2.12}) and (\ref{eq2.13})
together with the constraints (\ref{eq3.6}) lead to a solution 
with $N=1$:\footnote{
In \cite{Kodama}, 
Kodama has pointed out that the
exponential of the Chern-Simons action
is an exact solution of the Hamiltonian 
constraint in the holomorphic representation of 
the Ashtekar formalism \cite{Ashtekar}. 
Connections with our solution are unclear.
}
%
\begin{eqnarray}
S_{1}[g] = \int\! d^{D}\!x\,\sqrt{g(x)}\,
  \Big\{ \gamma_{0} + \gamma_{1}\,R(x)\Big\}\,,
\label{eq4.6}
\end{eqnarray}
%
where
%
\begin{eqnarray}
\gamma_{0} \!\!&=&\!\!
  \frac{4(D-1)^{2}(D-2)!\,(4\pi)^{D/2}}{(D^{2}-2)\phi^{(D)}(0)}
   \left( 2\Lambda + \frac{6D(D-1)\phi^{(D+2)}(0)}
    {(D+2)(D^{3}+10D^{2}-13D-34)\phi^{(D)}(0)} \right),\nonumber\\
\gamma_{1} \!\!&=&\!\!
  \,\frac{48(D-1)D!\,(4\pi)^{D/2}}
         {(D-2)(D^{3}+10D^{2}-13D-34)\phi^{(D)}(0)}\,.
\label{eq4.7}
\end{eqnarray}
%
Thus, we conclude that the renormalized Wheeler-DeWitt equation 
has a $z=2$ solution to leading order of the strong coupling
expansion.

The form of the leading solution (\ref{eq4.6}) is good news
for the renormalizability of the theory and our strong coupling
approximation.
In our formulation, the renormalizability requires that
all physical quantities must be independent of the arbitrary
function $\phi(\varepsilon)$ or $\phi^{(n)}(0)$.
Fortunately, this can be achieved, at least to leading order,
by absorbing $\phi^{(D)}(0)$ and $\phi^{(D+2)}(0)$ into the
redefinition of $G$ and $\Lambda$ because the leading order
wave functional becomes independent of $\phi$.
This implies that the combination $G^{2}(\mu)\,\mu^{D}$,
where $\mu$ is a mass parameter defined by $\phi^{(D)}(0)
\sim \mu^{D}$, should be independent of $\mu$.
This fact makes a physical meaning of our approximation clear.
Since the actual dimensionless expansion parameter is 
$1/[G^{2}(\mu)\mu^{2(D-1)}]$ and since it tends to zero as the
mass scale $\mu$ increases (if $D>2$), our strong coupling
expansion is thus expected to give a good approximation
scheme at high energies, as mentioned before.

Let us next try to construct solutions associated with 
higher Lifshitz points.
Suppose that $S^{\,\prime}_{1}[g]$ is another solution to (\ref{eq4.4}).
Then, $S^{\,\prime}_{1}[g] - S_{1}[g]$ has to satisfy
%
\begin{eqnarray}
\Delta_{R}(x)\,\Big( S^{\,\prime}_{1}[g] - S_{1}[g] \Big) = 0\,.
\label{eq4.8}
\end{eqnarray}
%
If $S^{\,\prime}_{1}[g]$ would correspond to a $z=2N$ solution,
$S^{\,\prime}_{1}[g] - S_{1}[g]$ should be expanded as\footnote{
We can add a gravitational Chern-Simons term discussed in
\cite{HoravaGR2} to (\ref{eq4.9}).
However, the conclusion given below will not change.
}
%
\begin{eqnarray}
S^{\,\prime}_{1}[g] - S_{1}[g]
  = \int\!d^{D}\!x\,\sqrt{g(x)}\,\sum_{a}\sum_{n=0}^{N}
    \delta^{a}_{n}\,{\cal O}^{a}_{2n}(x)\,.
\label{eq4.9}
\end{eqnarray}
%
It turns out that the action of $\Delta_{R}(x)$ on (\ref{eq4.9})
will lead to
%
\begin{eqnarray}
\Delta_{R}(x) \Big( S^{\,\prime}_{1}[g] - S_{1}[g] \Big)
  = \sqrt{g(x)}\,\sum_{a}\sum_{n=0}^{N}
    \delta^{\,\prime a}_{n}\,{\cal O}^{a}_{2n}(x)\,.
\label{eq4.10}
\end{eqnarray}
%
Since ${\cal O}^{a}_{2n}$'s are independent each other, all
the coefficients $\delta^{\,\prime a}_{n}$ have to vanish 
to be a solution to (\ref{eq4.8}).
Note that the number of the parameters $\{\delta^{a}_{n}\}$ 
is, in general, less than that of $\{ \delta^{\,\prime a}_{n}\}$ because
if ${\cal O}^{a}_{2n}$ is written as a total derivative
like $\nabla_{i}\nabla^{i}R$, it does not contribute to
(\ref{eq4.9}) but can appear in (\ref{eq4.10}).
Therefore, all the coefficients $\delta^{a}_{n}$ in (\ref{eq4.9})
should be trivial 
as long as there is no accidental degeneracy
in the relations between $\delta^{a}_{n}$ and 
$\delta^{\,\prime a}_{n}$.
For $D=3$ with $N=2$, we can explicitly verify that there is 
no nontrivial solution $S^{\,\prime}_{1}[g]$.
We thus conclude that the solution (\ref{eq4.6}) is
unique to leading order in the strong coupling expansion
in a class of integrals of local functions given 
in (\ref{eq4.5}).\footnote{
This conclusion does not mean that there are no nonlocal
solutions which include infinitely many higher derivatives.
}
This result shows that the quantum Einstein gravity
in our Wheeler-DeWitt formulation 
possesses a unique $z=2$ Lifshitz point
in the strong coupling limit.

Before closing this section, we would like to make a few comments
in order.
First, it should be stressed that the strong coupling expansion
is not a derivative expansion because higher derivative terms,
like $R^{m}\ (m > 1)$, could appear on the right-hand-side of
(\ref{eq4.6}) but it happens that their coefficients are zero
as a solution to the equation (\ref{eq4.4}).
Since our results show that in the strong coupling limit the
$D\!+\!1$-dimensional quantum Einstein gravity reduces to the 
$D$-dimensional Einstein gravity, one might expect that by the
inverse Wick rotation the $(D\!-\!1)+1$-dimensional theory could 
further reduce to $(D\!-\!1)$-dimensional one.
This is not, however, the case because our formulation can
apply only for the case of $D=$ odd due to the anomaly in 
(\ref{eq3.2}).
The final comment is that we need the \lq\lq cosmological" term $\gamma_{0}$
in (\ref{eq4.6}) in order for (\ref{eq4.6}) to become a solution
to (\ref{eq4.4}) even if $\Lambda=0$.

\vspace{5mm}
%
%
%
\section{Conclusions}
In this paper, we have solved the renormalized Wheeler-DeWitt
equation in $D+1$ dimensions in the strong coupling limit
to test Ho\v{r}ava's idea that quantum gravity
at short distances will be described by a nonrelativistic
theory with $z>1$.
Our results indicate that the dimensional reduction 
\cite{Parisi-Wu} from $D+1$ to $D$ dimensions for $D=$ odd
occurs in the strong coupling limit and the quantum Einstein gravity
has a $z=2$ Lifshitz point but no other higher Lifshitz points.
Although we have not found a $z=3$ solution \cite{HoravaGR2,HoravaGR3}, 
which will lead to a power-counting renormalizable
quantum gravity theory in $3+1$ dimensions,
we may still have a chance to get a finite theory because
the 3-dimensional Einstein gravity has no local excitation
and can be described by a topological field 
theory \cite{3-d gravity}.

Our analysis can also be applied to the functional 
Schr\"{o}dinger equation for the Yang-Mills theory.
In \cite{Greensite, Mansfield, SakamotoYM}, it has been
suggested that the dimensional reduction from $3+1$ to $3$
dimensions occurs in the infrared region and a vacuum
wave functional for the $3+1$-dimensional Yang-Mills
theory is given by the exponential of the $3$-dimensional 
Yang-Mills action.
This result now has a new interpretation that the Yang-Mills
theory can be described by a $z=2$ nonrelativistic theory
at low energies.\footnote{
This happens in the infrared region,
although Ho\v{r}ava \cite{HoravaYM} supposed that this 
situation will occur at high energies. 
}
It would be of great interest to investigate the
Yang-Mills theory at low energies from a new perspective.

\vspace{5mm}
%
%
%
\begin{center}
{\bf Acknowledgments}
\end{center}
This work is supported in part by a Grant-in-Aid for Scientific 
Research (No.18540275) from the Japanese Ministry of Education, 
Science, Sports and Culture. 
The author would like to thank M. Kato for bringing our attention
to the papers \cite{HoravaGR1,HoravaGR2}
and also thank T. Horiguchi and K. Maeda for collaboration at an
early stage of this work.
%
%
%
%

\vspace{10mm}
%
%
%

%
%
%
%
%
\end{document}